\documentclass[a4paper,10pt]{article}
\usepackage{pos}
\usepackage{float}
\usepackage{siunitx}
\usepackage{booktabs}
\usepackage{mathtools}
\usepackage{physics}
\usepackage{bbm}
\usepackage[noabbrev, nameinlink]{cleveref}
\creflabelformat{equation}{#2\textup{#1}#3}
\usepackage{csquotes}
\usepackage{layouts}
\usepackage{sidecap}
\sidecaptionvpos{figure}{m}
\sidecaptionvpos{table}{m}
\title{Topology changing update algorithms\\ for SU(3) gauge theory}
\ShortTitle{Topology changing update algorithms for SU(3) gauge theory}

\author*{Timo Eichhorn}
\author{Christian Hoelbling}
\author{Philip Rouenhoff}
\author{Lukas Varnhorst}

\affiliation{Department of Physics, University of Wuppertal, Gaußstraße 20, D-42119 Wuppertal, Germany}

\emailAdd{timo.eichhorn@protonmail.com}

\abstract{At fine lattice spacings, lattice simulations are plagued by slow (topological) modes that give rise to large autocorrelation times. These, in turn, lead to statistical and systematic errors that are difficult to estimate. We study the problem and possible algorithmic solutions in 4-dimensional SU(3) gauge theory, with special focus on instanton updates and Metadynamics.}

\FullConference{%
The 39th International Symposium on Lattice Field Theory,\\
8th-13th August, 2022,\\
Rheinische Friedrich-Wilhelms-Universität Bonn, Bonn, Germany
}

\begin{document}
\maketitle
\newcommand\thefont{\expandafter\string\the\font}

\section{Introduction}
Current state of the art calculations in lattice field theory rely on Monte Carlo simulations that sample the relevant probability distributions via Markov chains, as the underlying probability distributions are too complicated to be sampled directly, although there are ongoing efforts to leverage normalizing flows to directly sample these probability distributions \cite{Albergo:2019eim, Kanwar:2020xzo, Boyda:2020hsi, Albergo:2021bna, Hackett:2021idh, Foreman:2021ljl, Finkenrath:2022ogg, Albergo:2022qfi, Abbott:2022zhs, Abbott:2022hkm}. Subsequent states in Markov chains are correlated, which needs to be taken into account during the analysis and estimation of statistical and systematic uncertainties. These correlations can be especially problematic in the vicinity of critical points, where many autocorrelation times associated with observables measured on the generated configurations diverge.

Generally, the integrated autocorrelation times near second-order phase transitions are expected to be described by $\tau_{\textrm{int}} \propto \xi^{z}$, where $\xi$ is the correlation length and $z$ the dynamical critical exponent depending on the update algorithm(s) being used. For Langevin-type algorithms, the dynamical critical exponent can be shown to be $z = 2$ \cite{Baulieu:1999wz}, and for local update algorithms, a similar behaviour is expected due to their diffusive nature. Empirically, the Hybrid Monte Carlo (HMC) algorithm has also been found to behave similarly to Langevin-type algorithms for QCD and SU(3) gauge theory, but the scaling behaviour of the HMC is not predictable for interacting theories due to its non-renormalizability \cite{Luscher:2011qa}.

For topologically nontrivial gauge theories, however, the autocorrelation times below certain lattice spacings are dominated by topological modes stemming from high action barriers separating the topological sectors which emerge towards the continuum limit. This problem, commonly known as topological freezing, has been observed in previous studies of both QCD and other lattice gauge theories \cite{Alles:1996vn, DelDebbio:2002xa, DelDebbio:2004xh, Schaefer:2010hu}. There, the rise in autocorrelation times associated with the topological charge was found to be compatible with a dynamical critical exponent $z = 5$, or an exponential behaviour commonly found near first-order phase transitions. Although this seems to contradict the previous statements concerning the scaling behaviour of the algorithms, this paradox is resolved by noting that the topology changing transitions may be interpreted as non-perturbative lattice artifacts absent from alternative formulations of QCD using open boundaries \cite{Luscher:2011kk}.

Of course, topological freezing is highly problematic if one is interested in calculating observables closely related to the topological charge such as the topological susceptibility or the $\eta'$ mass.
But even for other non-topological observables, slow topological modes may be problematic: For instance, while (smeared) Wilson loops show indications of decoupling from the slow topological modes \cite{Schaefer:2009xx, Schaefer:2010hu}, it is generally extremely difficult to estimate to which degree observables decouple, which introduces unknown statistic and systematic errors.

Here, we study the application of Metadynamics \cite{Laio_2002, Laio:2015era} and instanton updates \cite{FUCITO1984230, Smit:1987fq, Dilger:1992yn, Dilger:1994ma, Durr:2012te} to 4-dimensional SU(3) gauge theory, based upon our investigation in 2-dimensional U(1) gauge theory \cite{Eichhorn:2021ccz, Hoelbling:2022wru}.
\section{Conventional update algorithms}
To illustrate the effects of topological freezing and establish a baseline to compare our results to, we generated configurations for the same physical volume with the Wilson gauge action at different lattice spacings using four update schemes: A single heat bath sweep \cite{Creutz:1980zw, Fabricius:1984wp, Kennedy:1985nu} (1 HB), five heat bath sweeps (5 HB), a combined update of one heat bath sweep followed by four overrelaxation sweeps \cite{Adler:1981sn, Creutz:1987xi, Brown:1987rra} (1 HB + 4 OR), and a single HMC update \cite{Duane:1987de} with unit length trajectory (1 HMC). Following Cabibbo and Marinari, the heat bath and overrelaxation updates were applied to three different diagonal SU(2) subgroups \cite{Cabibbo:1982zn}. The equations of motion during the HMC were integrated using an Omelyan-Mryglod-Folk fourth-order minimum norm integrator \cite{OMELYAN2003272} with a stepsize of $\epsilon = 0.2$, yielding acceptance rates around $99\%$ for all parameters considered here. Our simulation parameters are summarized in \Cref{tab:simulated_parameters}.
\begin{SCtable}[][h]
    \centering
    \begin{tabular}{cccc}
        \bottomrule
         $\beta$ & $L/a$ & $a$ [fm] & $N_{\textrm{conf}}$ \\ \toprule
         5.8980 & $10$ & $0.1097$ & $100000$ \\
         6.0000 & $12$ & $0.0914$ & $100000$ \\
         6.0938 & $14$ & $0.0783$ & $100000$ \\
         6.1802 & $16$ & $0.0686$ & $100000$ \\
         6.2602 & $18$ & $0.0610$ & $100000$ \\
         6.3344 & $20$ & $0.0549$ & $100000$ \\
         6.4035 & $22$ & $0.0481$ & $100000$ \\ \toprule
    \end{tabular}
    \caption{A summary of the simulation parameters for the runs using conventional update algorithms. The scale was set via the rational fit from \cite{Durr:2006ky}, which in turn used data from \cite{Necco:2001xg}.}
    \label{tab:simulated_parameters}
\end{SCtable}

The topological charge is defined through a gluonic definition, i.e., a sum over a topological charge density constructed out of a suitably defined lattice field strength tensor. We employ the usual clover definition with lattice artifacts of $\mathcal{O}(a^2)$; as an additional crosscheck we consider the plaquette-based definition which also suffers from $\mathcal{O}(a^2)$ artifacts. In the following, $Q_{c}$ and $Q_{p}$ refer to the clover and plaquette definitions respectively. The operators are applied to the gauge fields after various sweeps of stout smearing \cite{Morningstar:2003gk} with a smearing parameter of $\rho = 0.12$. Other observables we consider are $n \times n$ Wilson loops and the Wilson gauge action at various smearing levels, denoted by $\mathcal{W}_{n}$ and $S_w$ respectively. All autocorrelation times and their uncertainties are estimated following the procedure described in \cite{Wolff:2003sm}. To allow for a fair comparison, the relative times required for one update using the different update schemes are listed in \Cref{tab:update_times}.
\begin{SCtable}[1.5][h]
    \centering
    \begin{tabular}{cc}
        \bottomrule
         Update scheme & Relative time \\ \toprule
         1 HB & 1.00 \\
         5 HB & 4.99 \\
         1 HB + 4 OR & 2.02 \\
         1 HMC & 6.98 \\ \toprule
    \end{tabular}
    \caption{The relative performances of the four different update schemes used here, based on measurements on the $22^4$ lattice. Note that the heat bath update is $\beta$-dependent: With higher $\beta$, the heat bath update becomes slightly faster due to the fact that on average, less iterations are required to sample the correct probability distribution (cf. \cite{Fabricius:1984wp, Kennedy:1985nu}).}
    \label{tab:update_times}
\end{SCtable}

\begin{figure}[h]
    \centering
    \includegraphics[width=.495\textwidth]{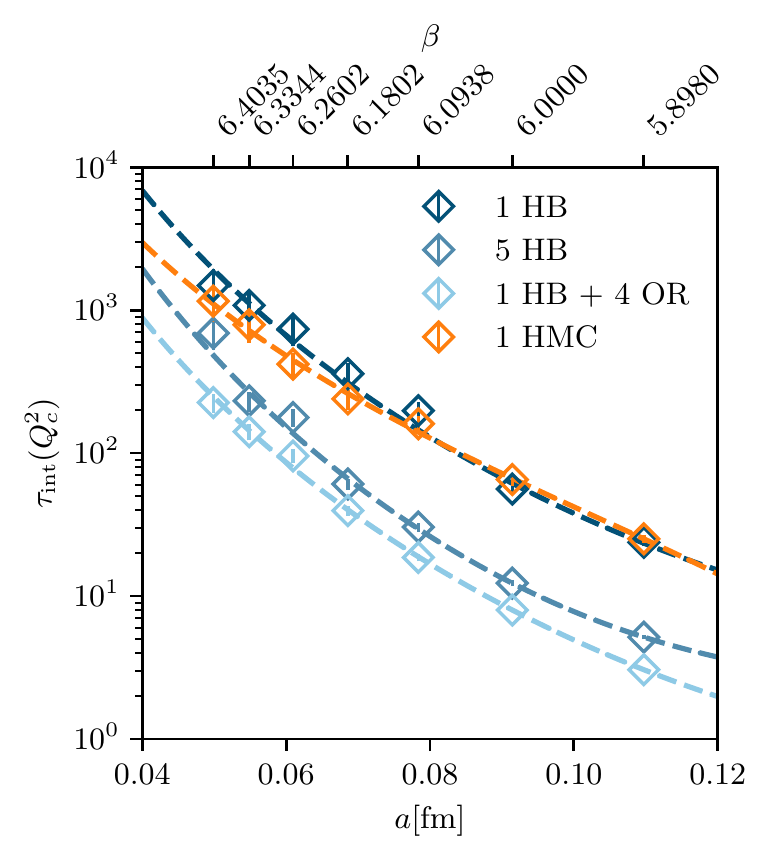}
    \includegraphics[width=.495\textwidth]{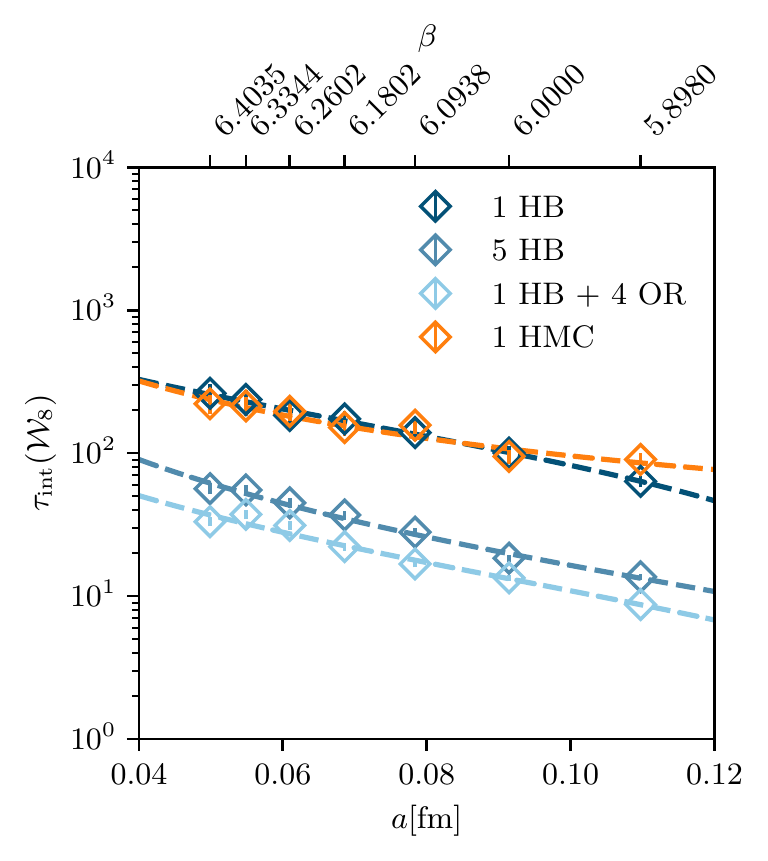}
    \caption{Scaling of the integrated autocorrelation times of $Q_c^2$ (left) and $\mathcal{W}_8$ (right) with the lattice spacing $a$. Both observables were measured after 31 stout smearing steps for four different update schemes. In the parameter range studied here, the data for $Q_c^2$ seems to be compatible with a power law behaviour with exponent $z \approx 5$ whereas the data for $\mathcal{W}_8$ seems to scale with a much smaller exponent. The dynamical critical exponents obtained from power law fits can be found in \Cref{tab:tau_int_fit_parameters}.}
    \label{fig:tau_int_scaling}
\end{figure}

\Cref{fig:tau_int_scaling} shows the scaling of the integrated autocorrelation times of $Q_c^2$ and $\mathcal{W}_8$ measured after 31 stout smearing steps. For $Q_c^2$ and $Q_p^2$, there is a drastic increase of the integrated autocorrelation times towards the continuum that is compatible with a power law behaviour with an exponent $z = 5$, or an exponential behaviour, both of which are consistent with previous findings \cite{DelDebbio:2002xa, Schaefer:2009xx, Schaefer:2010hu}. Here, we only show the power law fit to $Q_c^2$; the autocorrelation times for $Q_p^2$ agree with those for $Q_c^2$ within uncertainties, and the exponential fit is very similar to the power law fit within the range studied here. For smeared Wilson loops, the increase is much milder and can be described by a power law with an exponent $z \approx$ \numrange{1}{2}. From this comparison it is evident that the topological modes only begin to dominate below $a \approx \SI{0.08}{\femto\meter}$, whereas for coarser lattice spacings, smeared Wilson loops display larger autocorrelation times.
\begin{table}[h]
    \centering
    \begin{tabular}{cccccc}
        \bottomrule
         Update scheme & $z(Q_{c}^{2})$ & $z(S_w)$ & $z(\mathcal{W}_{2})$ & $z(\mathcal{W}_{4})$ & $z(\mathcal{W}_{8})$ \\ \toprule
         1 HB & \num{5.77 \pm 0.57} & \num{0.66 \pm 0.37} & \num{0.65 \pm 0.38} & \num{0.65 \pm 0.36} & \num{0.56 \pm 0.35} \\
         5 HB & \num{6.33 \pm 0.38} & \num{1.31 \pm 0.48} & \num{1.34 \pm 0.49} & \num{1.43 \pm 0.50} & \num{1.59 \pm 0.48} \\
         1 HB + 4 OR & \num{5.75 \pm 0.20} & \num{0.92 \pm 0.72} & \num{0.93 \pm 0.73} & \num{1.00 \pm 0.73} & \num{1.12 \pm 0.64} \\
         1 HMC & \num{4.46 \pm 0.32} & \num{0.33 \pm 0.87} & \num{0.42 \pm 0.91} & \num{0.8 \pm 1} & \num{1.4 \pm 1.3} \\ \toprule
    \end{tabular}
    \caption{The dynamical critical exponents obtained from power law fits to the integrated autocorrelation times of $Q_{c}^{2}$, $S_w$, and Wilson loops of different sizes after 31 stout smearing steps. Notably, the dynamical critical exponent associated with $Q_c^2$ is much larger than those associated with the smeared action or smeared Wilson loops of different sizes.}
    \label{tab:tau_int_fit_parameters}
\end{table}

A single heat bath update and a unit length HMC update seem to perform similarly in terms of autocorrelation times, but the HMC is more computationally expensive by a factor of about $7$. However, it is possible to adopt a local, more efficient version of the HMC for pure gauge theory \cite{Marenzoni:1993im}, which reduces the number of staple calculations and allows for larger integration steps. As we are ultimately interested in simulations including dynamical fermions, we did not test the local HMC. The combination of heat bath and overrelaxation updates appears to lead to the lowest autocorrelation times out of all update schemes studied here, with the autocorrelation times being about a factor of 2 smaller than those observed using the 5 HB update scheme, while also being faster by about a factor 2. Still, the inclusion of overrelaxation updates does not seem to improve the scaling behaviour.
\section{Metadynamics}
Metadynamics modifies the action of the simulated system with a time-dependent bias potential, thereby encouraging the system to visit previously unexplored regions of the configuration space. This is accomplished by introducing a set of collective variables (CVs) which effectively span a low-dimensional projection of the original configuration space. As the potential reaches equilibrium, it converges towards the negative free energy/action of the system \cite{PhysRevLett.96.090601} (up to a constant offset), and the resulting probability distribution will be approximately flat with respect to the CVs. Therefore, Metadynamics can be used to overcome high action barriers as long as the CVs are able to properly resolve the action landscape. In the end, observables need to be reweighted to obtain expectation values with respect to the original probability distribution. For $\textrm{CP}^{N - 1}$ models, it has already been demonstrated that Metadynamics can be used to alleviate topological freezing \cite{Laio:2015era}, and exploratory studies of Metadynamics \cite{Bonati:2017nhe} and related methods \cite{Bonati:2018blm} applied to QCD also seem promising.

In the context of topological freezing, an obvious candidate for the CV is a non-integer observable correlated with the topological charge, but the choice is somewhat ambiguous: Directly using a gluonic definition leads to an observable that is effectively uncorrelated with the topological charge, since the signal is contaminated by UV noise, so some degree of smoothing is necessary. Smoothing the gauge fields too much, however, will lead to the values being peaked too strongly around integers, which will eventually drive the system to very coarse configurations with high actions. Furthermore, the computational overhead introduced by the additional smoothing is significant in the case of pure gauge theory, which adds another incentive to reduce the number of smoothing steps. Still, even for a few smearing steps, our CV effectively introduces an additional, less local term to the action, which makes local update algorithms impractical to use. Instead, we use the HMC, which restricts our smoothing method to be differentiable with respect to the gauge fields. Therefore, we chose our CV to be given by the clover-based definition applied to fields after $n_s =$ \numrange{5}{10} iterations of stout smearing with a smearing parameter of $\rho = 0.12$. To distinguish between the CV and the observable used during measurements, we denote the CV by $Q_{\textrm{meta}}$. The additional force term induced by the metapotential is given by the derivative of the metapotential with respect to the gauge field:
\begin{equation}
    F_{\textrm{meta}} = - \frac{\partial V_{\textrm{meta}}}{\partial U} =
    - \frac{\partial V_{\textrm{meta}}}{\partial Q_{\textrm{meta}}} \frac{\partial Q_{\textrm{meta}}}{\partial U^{(n_s)}} \frac{\partial U^{(n_s)}}{\partial U}
    \label{eq:metadynamics_force}
\end{equation}
The derivative of the clover-based charge with respect to the maximally smeared fields $U^{(n_s)}$ consists of a sum of staples with clover insertions, and the last term in \cref{eq:metadynamics_force} involves the derivative of the maximally smeared field with respect to the unsmeared field $U$, which corresponds to the usual stout force recursion appearing in fat-link stout-smeared actions \cite{Morningstar:2003gk, Durr:2010aw}.

\begin{figure}[H]
    \centering
    \includegraphics[width=\textwidth]{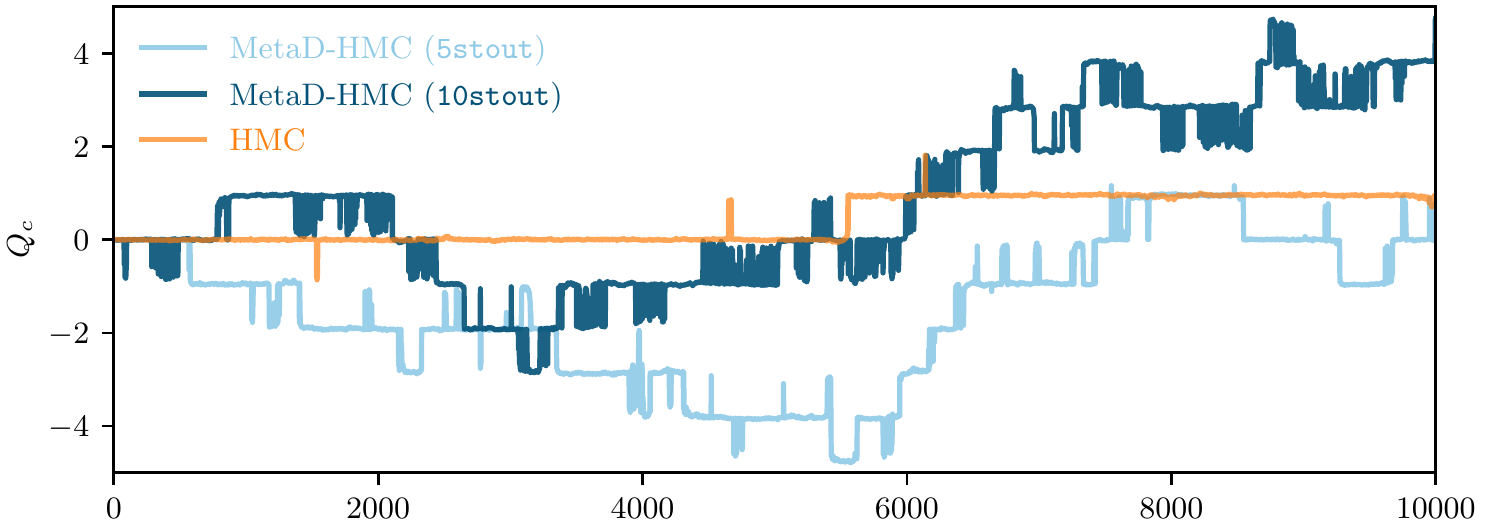}
    \caption{Comparison of the time series of the topological charge between a normal HMC run and two MetaD-HMC runs with different smearing levels. Both MetaD-HMC runs tunnel much more frequently between the topological sectors than the standard HMC run.}
    \label{fig:timeseries_comparison_22}
\end{figure}

\Cref{fig:timeseries_comparison_22} shows a comparison between the time series of the topological charge obtained using the HMC and the HMC with Metadynamics (MetaD-HMC) with five and ten stout smearing steps for a $22^4$ lattice with $\beta = 6.4035$.
Compared to the HMC which has difficulties tunneling between topological sectors, the MetaD-HMC tunnels between sectors much more frequently. However, if the number of smearing steps used to define the CV is set too high, the topological charge undergoes frequent fluctuations without settling in one sector, possibly due to the system being driven to very coarse configurations by the metapotential. As seen in \Cref{fig:timeseries_comparison_22}, these fluctuations are less prominent if five instead of ten stout smearing steps are used, while the tunneling behaviour is not negatively affected, so a proper choice of parameters seems crucial.

Overall, the more frequent tunneling comes at the cost of a considerable computational overhead: Depending on the number of smearing steps, the MetaD-HMC is around 20 to 30 times slower than the conventional HMC, with most of the time spent during the smearing of the fields and the stout force recursion. Thus, it is likely that the combination of heat bath and overrelaxation updates is more efficient at decorrelating the topological charge for the current setup. We expect the overhead to be less severe at finer lattice spacings, where the number of required smearing steps to define an appropriate CV should decrease. Other changes that might reduce the number of required smearing steps could be the use of an improved action and topological charge operator.

A potential issue of the method is the relatively long time required for the metapotential to equilibrate, especially when considering larger volumes where the range of relevant topological sectors is much broader. For instance, in the runs shown in \Cref{fig:timeseries_comparison_22}, the metapotentials were still not in equilibrium after $10000$ updates. We are currently investigating methods to accelerate or completely circumvent the buildup of the metapotential.
\section{Instanton updates}
\begin{SCfigure}[][h]
    \centering
    \includegraphics[width=.75\textwidth]{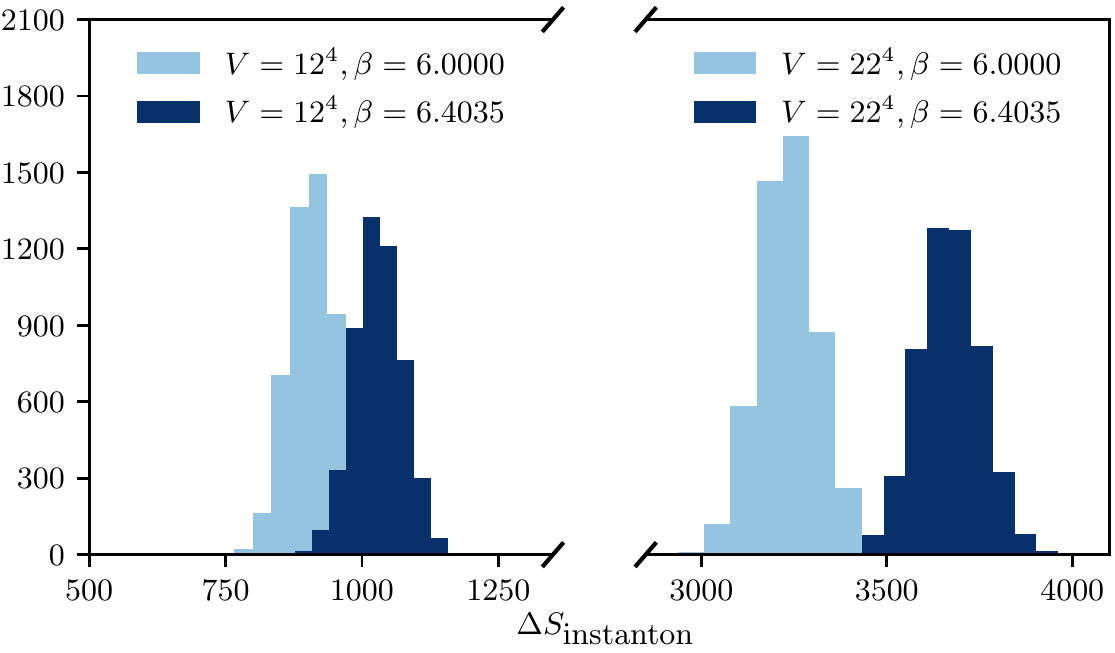}
    \caption{Histogram of the action differences for the instanton update for different parameters. The average action difference increases with the volume and the inverse coupling $\beta$, but even for the $12^4$ lattice at $\beta = 6.0$, the average action difference is too high to achieve reasonable acceptance rates.}
    \label{fig:deltaH_instanton}
\end{SCfigure}

\begin{figure}[h]
    \centering
    \includegraphics[width=.495\textwidth]{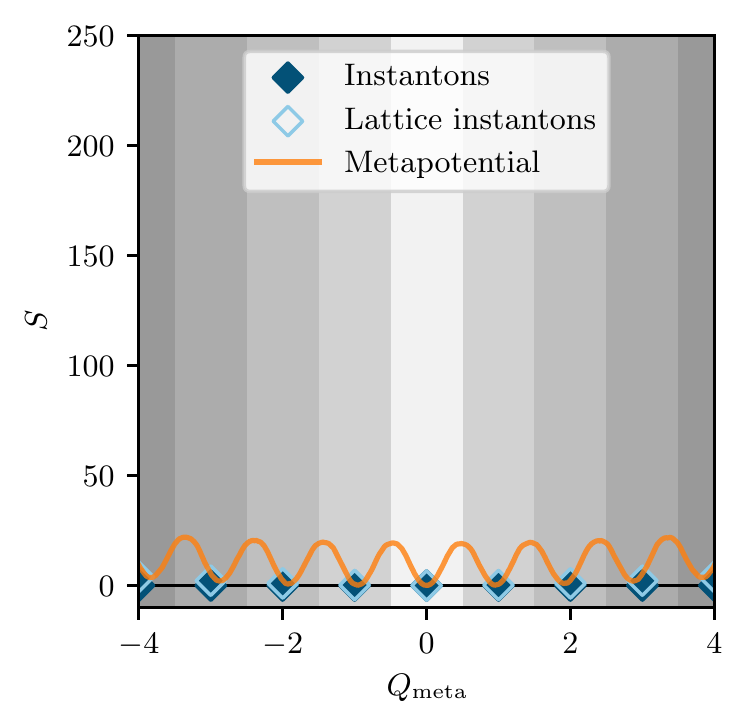}
    \includegraphics[width=.495\textwidth]{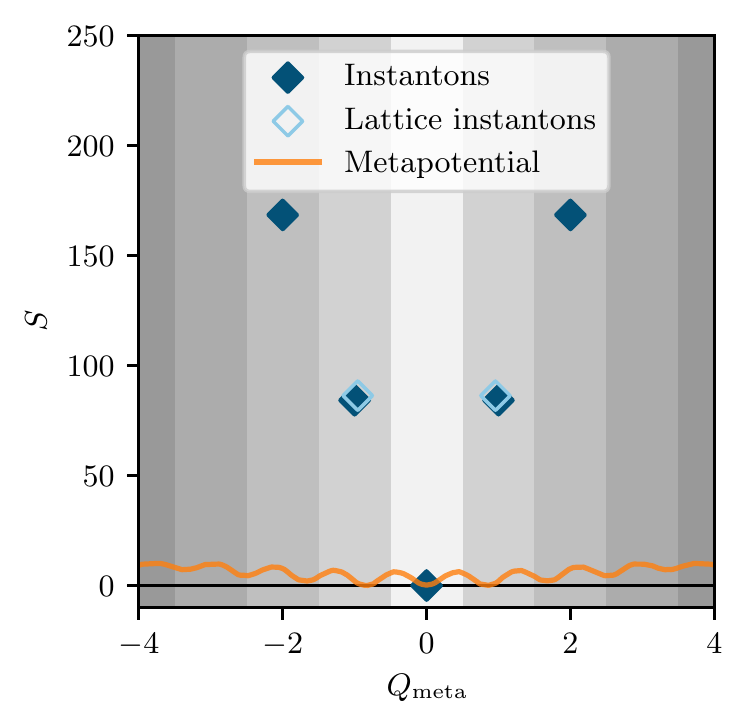}
    \caption{The metapotentials and instanton actions for 2-dimensional U(1) on a $32^2$ lattice at $\beta = 12.8$ (left) and 4-dimensional SU(3) on a $22^4$ lattice at $\beta = 6.4035$ (right). The points labeled \enquote{Instantons} refer to the continuum infinite volume action. For 2-dimensional U(1), the minima of the metapotential match the instanton actions well, whereas for 4-dimensional SU(3), the action difference between instantons is much higher than the action difference between the adjacent metapotential minima.}
    \label{fig:instanton_metapotentials}
\end{figure}

Instanton updates propose new configurations by link-wise multiplication of the previous configuration with an instanton configuration of charge $Q_{i} = \pm 1$ \cite{Belavin:1975fg}, where the sign is randomly chosen with equal probability. The newly proposed configuration is then subject to a conventional Metropolis accept-reject step. When combined with an ergodic update algorithm, instanton updates work well in 2-dimensional U(1) theory and the Schwinger model, since they avoid the high action regions of phase space between topological sectors by directly jumping from sector to sector. With generally high acceptance rates that are only weakly dependent on the gauge coupling \cite{Eichhorn:2021ccz}, the topological charge is effectively decorrelated after a few updates at most. To generalize the algorithm to 4-dimensional SU(3) theory, we first consider the construction of lattice instantons. For 4-dimensional SU(2), the gauge potential of an instanton centered around $z$ in singular gauge reads:
\begin{equation}
    A_{\mu}^{\textrm{SU(2)}} = \eta_{a \mu \nu} \frac{\rho^2 (x_{\nu} - z_{\nu}) \tau_{a}}{(x - z)^2 ((x - z)^2 + \rho^2)}
    \label{eq:BPST_instanton}
\end{equation}
Here, $\eta_{a \mu \nu}$ is the 't Hooft symbol, $\rho$ a scale parameter of the instanton, and $\tau_{a}$ corresponds to $-i \sigma_{a}$, where $\sigma_{a}$ denotes the $a$-th Pauli matrix. An anti-instanton can be obtained by replacing the 't Hooft symbol with the anti-'t Hooft symbol $\Bar{\eta}_{a \mu \nu}$. Note that this only holds if we define the topological charge as:
\begin{equation}
    Q = \frac{1}{32 \pi^2} \int d^4x \tr[\epsilon_{\alpha \beta \gamma \delta} F_{\alpha \beta} F_{\gamma \delta}]
\end{equation}
If the topological charge is defined with the opposite sign, \cref{eq:BPST_instanton} already describes an anti-instanton.
The extension to SU($N$) is performed by embedding the SU(2) solution into SU($N$). Exponentiating the gauge potential along a link connecting two lattice sites, we get:
\begin{equation}
    U_{\mu}(n; z, \rho) =  \cos(\lambda_{\mu}) + i \sigma_{a} \eta_{a \mu \nu} \sin(\lambda_\mu) \frac{(x_\nu - z_\nu)}{\sqrt{(x - z)^2 - (x_\mu - z_\mu)^2}}
\end{equation}
For more details of the derivation and the definition of $\lambda_{\mu}$, we refer to Appendix B in \cite{Jahn:2019nmd}. In practice, the instanton is centered around the middle of a hypercube in order to avoid the singularity of the gauge potential at the origin. For simplicity, this hypercube was always chosen to be the central hypercube of the lattice. The configurations constructed in this way are discontinuous at the periodic boundaries of the lattice, which increases the action of the configuration compared to the infinite volume case. These discontinuities can be dealt with by smearing the configurations. For the setup studied here, the smallest instanton size still stable under this smearing procedure was $\rho_{\textrm{min}} = 5$, although it is possible that smearing with respect to a different (improved) action or varying the smearing parameter with the distance to the instanton center as done in \cite{Jahn:2019nmd} might allow the construction of smaller stable lattice instantons. We found that after applying 150 stout smearing steps, the action of the lattice instanton configuration would reach a plateau at a value around \SI{10}{\percent} above the continuum infinite volume action, while the topological charge was a few percent below the exact value. A visual comparison of these quantities can be found in \Cref{fig:instanton_metapotentials}. Using these approximate lattice instantons, we measured the action differences upon multiplying them onto thermalized configurations, which were generated using the 1 HB + 4 OR update scheme. The results are shown in the histograms in \Cref{fig:deltaH_instanton}.

The action differences after multiplication with an (anti-)instanton are much too high to achieve reasonable acceptance rates: They would have to be between two to three orders of magnitude smaller to achieve acceptance rates close to \SI{1}{\percent}. While we have not tested the multiplication with instantons of different sizes, we do not expect there to be a choice of parameters that will sufficiently decrease the action differences.

It is potentially more interesting to understand the cause of the high action differences, especially in light of the much smaller action differences seen when applying instanton updates in 2-dimensional U(1) theory and the Schwinger model. Some more insight can be gained by comparing the differences between instanton actions with the metapotentials, which allow one to estimate the differences between the average actions in the different topological sectors: For 2-dimensional U(1), the action of a lattice (anti-)instanton with topological charge Q is given by
\begin{equation}
    S = \beta V \left( 1 - \cos(\frac{2 \pi}{V} \abs{Q}) \right) \xrightarrow[V \rightarrow \infty]{} 0,
\end{equation}
whereas for 4-dimensional SU(3), the action of an (anti-)instanton with topological charge Q is given by \cite{Bogomolny:1975de}
\begin{equation}
    S = \frac{8 \pi^2}{g^2} \abs{Q} = \frac{\beta}{6} 8 \pi^2 \abs{Q}.
    \label{eq:Bogomolnyi_bound}
\end{equation}
\Cref{fig:instanton_metapotentials} shows that in 2-dimensional U(1), the finite volume corrections to the instanton action match the minima of the metapotential quite well. In 4-dimensional SU(3) on the other hand, these effects are clearly subleading because they are an order of magnitude smaller than the instanton action. The fact that the instanton action vanishes in 2-dimensional U(1) seems to be essential to obtain reasonable acceptance rates for such an update scheme. Configurations leading from one sector to another without large action jumps are therefore not expected to be similar to instanton configurations in 4-dimensional SU(3).
In the hope of increasing acceptance rates, which are extremely small due to the large action penalty associated with even an infinite volume instanton, we modified the instanton update by first evolving the fields along the Wilson flow \cite{Luscher:2009eq} in one time direction, then multiplying the fields with an instanton configuration, and finally flowing back in the opposite time direction. If we construct an update algorithm involving such a transformation, the probability $p$ during the accept-reject step now also includes the Jacobian matrix $\mathcal{J}$ of the field transformations:
\begin{equation}
    p = \exp(-\Delta S + \ln(\det(\mathcal{J})))
\end{equation}
For the Wilson flow, the logarithm of the Jacobian is proportional to the integral of the plaquettes of the flowed fields with respect to the flow time $t$:
\begin{equation}
    \ln(\det(\mathcal{J})) = - \frac{16}{3} \int\limits_{0}^{t} dt' \mathcal{W}_{1}(U_{t'})
\end{equation}
Here, $U_{t'}$ denotes the fields at flow time $t'$. We varied the flow time, and tried evolving the fields to positive as well as negative flow times before the multiplication with an instanton. Unfortunately, all of our attempts lead to even lower acceptance probabilities than the original version of the instanton update. For some parameters, the lower acceptance rates were due to the Jacobian, while for others, an increased action difference compared to the unmodified instanton update was responsible.
\section{Conclusion}
In these proceedings, we studied the problem of topological freezing in 4-dimensional SU(3) gauge theory, and explored the application of Metadynamics and instanton updates as possible solutions. We found the increase in autocorrelation times towards the continuum limit to be consistent with previous results \cite{DelDebbio:2002xa, Schaefer:2010hu}: While the integrated autocorrelation times associated to observables directly related to the topological charge increase drastically towards the continuum (compatible with a dynamical critical exponent $z = 5$ or an exponential increase), other autocorrelation times associated to non-topological observables such as smeared Wilson loops show a much milder increase. As expected, out of the conventional update algorithms studied here, a combination of heat bath and overrelaxation updates seems to lead to the lowest autocorrelation times.

Metadynamics is successful at alleviating the critical slowing down of topological modes, but for the simulation of pure gauge theory, the overhead introduced by the stout smearing during each momentum update is considerable. Depending on the number of smearing steps used to define the CV, the MetaD-HMC was between 20 to 30 times slower than a conventional HMC. However, the number of required smearing steps decreases for finer lattice spacings, and for QCD simulations with dynamical fermions, the relative overhead is expected to be much smaller than in the case of pure gauge theory. Additionally, it is conceivable that the overhead can be somewhat reduced by a better choice of parameters. A potential issue is that the reweighting required to obtain expectation values with respect to the correct probability distribution decreases the effective sample size \cite{Hoelbling:2022wru}. Parallel tempering with respect to open boundaries might be an alternative that does not suffer from this drawback \cite{Hasenbusch:2017unr, Bonanno:2020hht}. In the future, we plan to study Metadynamics with improved actions, where the freezing is much more severe at equivalent lattice spacings, and the expected computational overhead compared to standard update algorithms smaller.

Instanton updates suffer from high action differences in 4-dimensional SU(3) compared to 2-dimensional U(1) or the Schwinger model, which renders them unusable in practice. This can be attributed to the different infinite volume behaviours of the instanton actions: In 2-dimensional U(1), the infinite volume instanton action vanishes, and the finite volume corrections are small and comparable in size to the average action differences between topological sectors. In 4-dimensional SU(3), on the other hand, the infinite volume instanton action is substantially larger than the average action differences between topological sectors. A modified version combining the Wilson flow with the multiplication of an instanton configuration was affected by similar problems. While it is obviously possible to find configurations that, when multiplied with a configuration in one topological sector, lead to another topological sector without large action differences, we do not expect these configurations to be similar to instantons in 4-dimensional SU(3).

\acknowledgments
We gratefully acknowledge helpful discussions with Szabolcs Borsanyi, Stephan Dürr, Fabian Frech, Jana Günther, and Kalman Szabo. Calculations were performed on a local PC cluster at the University of Wuppertal.
% \newpage
\bibliographystyle{JHEP}
\setlength{\bibsep}{0pt plus 0.9ex}
\footnotesize
\bibliography{literature.bib}
\end{document}